\documentclass[aps,prb,amsmath,twocolumn,amssymb,floatfixng,showpacs,
superscriptaddress,footinbib]{revtex4-1}
\pdfoutput=1
\usepackage[dvips]{graphics}
\usepackage{bm}
\usepackage{float}
\usepackage{epsfig}
\usepackage{enumerate}
\usepackage{amsmath}
\usepackage{color}
\usepackage{braket}
\usepackage{graphicx}
\usepackage{float}
\usepackage[colorlinks=true,linktoc=page,linkcolor=magenta,citecolor=blue]{hyperref}

\newcommand{\etal}{{\it et al.~}}
\newcommand{\ie}{{\it i.e.,\,\,}}
\newcommand{\eg}{{\it e.g.\,\,}}
\newcommand{\ua}{{\uparrow }}
\newcommand{\da}{{\downarrow }}

\newcommand\bea{\begin{eqnarray}}
\newcommand\eea{\end{eqnarray}}
\newcommand\beq{\begin{equation}}  
\newcommand\eeq{\end{equation}}

\newcommand{\non}{\nonumber}  
\newcommand*{\citen}[1]{%
	\begingroup
	\romannumeral-`\x % remove space at the beginning of \setcitestyle
	\setcitestyle{numbers}%
	\cite{#1}%
	\endgroup   
}

\begin{document} 
\title{Higher Order Topological Insulator via Periodic Driving} 

\author{Arnob Kumar Ghosh}
\email{arnob@iopb.res.in}
\affiliation{Institute of Physics, Sachivalaya Marg, Bhubaneswar-751005, India}
\affiliation{Homi Bhabha National Institute, Training School Complex, Anushakti Nagar, Mumbai 400085, India}
\author{Ganesh C. Paul}
\email{ganeshpaul@iopb.res.in}
\affiliation{Institute of Physics, Sachivalaya Marg, Bhubaneswar-751005, India}
\affiliation{Homi Bhabha National Institute, Training School Complex, Anushakti Nagar, Mumbai 400085, India}
\author{Arijit Saha}
\email{arijit@iopb.res.in}
\affiliation{Institute of Physics, Sachivalaya Marg, Bhubaneswar-751005, India}
\affiliation{Homi Bhabha National Institute, Training School Complex, Anushakti Nagar, Mumbai 400085, India}

%~~~~~~~~~~~~~~~~~~~~~~~~~~~~~~~~~~~~~~~~~~~~~~~~~~~~~~~~~~~~~~~~~~~~~~~~~~~~~~~~~~~~~~~~~
\begin{abstract}
We theoretically investigate a periodically driven semimetal based on a square lattice. The possibility of engineering both Floquet Topological Insulator featuring Floquet edge states and Floquet higher order topological insulating phase, accommodating topological corner modes has been demonstrated starting from the semimetal phase, based on Floquet Hamiltonian picture. Topological phase transition takes place in the bulk quasi-energy spectrum with the variation of the drive amplitude where Chern number changes sign from $+1$ to $-1$. This can be attributed to broken time-reversal invariance ($\mathcal{T}$) due to circularly polarized light. When the discrete four-fold rotational symmetry ($\mathcal{C}_4$) is also broken by adding a Wilson mass term along with broken $\mathcal{T}$, higher order topological insulator (HOTI), hosting in-gap modes at all the corners, can be realized. The Floquet quadrupolar moment, calculated with the Floquet states, exhibits a quantized value of $ 0.5$ (modulo 1) identifying the HOTI phase. We also show the emergence of the {\it{dressed corner modes}} at quasi-energy $\omega/2$ (remnants of zero modes in the quasi-static 
high frequency limit), where $\omega$ is the driving frequency, in the intermediate frequency regime.
%that while for high frequency drive the Floquet corner modes remain pristine at zero energy,} \textcolor{blue}{the dressed corner modes} \textcolor{red}{arise at quasi-energy $\omega/2$, where $\omega$ is %the driving frequency, in the intermediate frequency regime.}  
\end{abstract}
%~~~~~~~~~~~~~~~~~~~~~~~~~~~~~~~~~~~~~~~~~~~~~~~~~~~~~~~~~~~~~~~~~~~~~~~~~~~~~~~~~~~~~~~~~

\maketitle

%-----------------------------
\section{Introduction}
%------------------------------
Investigation of topological phases of matter~\cite{hasan2010colloquium,qi2011topological,kane2005quantum,bernevig2006quantum,konig2007quantum,hsieh2008topological,
chen2009experimental} has been at the heart of quantum condensed matter physics for more than a decade from both theoretical and experimental perspective. A wide class of systems hosting topological phases have been discovered, including the $\mathcal{Z}_2$ topological insulator (TI)~\cite{fu2007topological,fu2011topological}, Weyl semimetal~\cite{wan2011topological,burkov2011weyl,armitage2018weyl}, Dirac semimetal~\cite{young2012dirac} and the topological superconductors~\cite{sato2017topological,paul2018spin}. The fascinating fact about TIs are the emergence of a robust $(d-1)$ dimensional boundary states from 
$d$ dimensional insulating bulk which is the outcome of bulk-boundary correspondence.

Very recently, the advent of the higher order topological insulators (HOTI)~\cite{benalcazar2017quantized,benalcazar2017electric,schindler2018higher,langbehn2017reflection,PhysRevB.92.085126} has attracted immense interest in the modern condensed matter physics community. In contrast to the prevailing topological insulators (TIs), in an $n$-th order TI both the $d$ dimensional bulk and $(d-1)$ dimensional boundary remain gapped, whereas, the $(d-n)$ dimensional boundary exhibit gapless modes; where $n$ is the order of the HOTI. In this language, conventional TIs are called first order TI. 
Thus, in three dimensions (3D), one can realize second order TI (SOTI) with gapless states located at the one dimensional ($1$D) hinges and third order TI with gapless zero dimensional ($0$D) corner states. 
Similarly, in two dimensions (2D), SOTI exhibits gapless 0D corner modes while the edges remain gapped. These $(d-n)$ dimensional boundary states emerge as quantization of $n$-th order electric multipole~\cite{benalcazar2017quantized,benalcazar2017electric} moment with evanescing lower moments. Hence, the materials we may have so far identified as trivial due to the absence of $(d-1)$ dimensional boundary states, may turn out to be HOTI~\cite{Ezawaphosphorene}. Following the increasing interest on HOTI, few intriguing models have been proposed so far based on Kagome lattice~\cite{Ezawakagome}, transition metal dichalcogenides ($\rm XTe_2$, X=Mo,W)~\cite{wang2018higher,ezawa2019second}, SnTe~\cite{schindler2018higher} etc. In 2D, SOTIs supporting zero dimensional corner modes have been experimentally realized in acoustic material based on kagome lattice~\cite{XueAcousticKagome}, photonic crystals~\cite{PhotonicChen,PhotonicXie}, ferromagnetic resonance~\cite{PhysRevResearch.1.032013} and electrical circuits~\cite{Peterson2018,Imhof2018} setup. 

Engineering of periodically driven Floquet topological insulators out of a trivial system is a field of interest of its own~\cite{lindner2011floquet,rudner2013anomalous,rudner2019floquet}. In this direction, 
realization of Floquet HOTI has been investigated from different perspectives in recent literature~\cite{tanaka2019appearance,szumniak2019hinge,luo2019higher,PhysRevB.100.115403,huang2018higher,hu2019higher,franca2018anomalous,rodriguez2019higher,PhysRevLett.123.016806,peng2019higher,PhysRevB.99.045441}. Among them, Floquet HOTI phase, hosting 
0D Floquet corner modes, has been studied on quantum spin hall system under a quantum quench~\cite{nag2019higher} with a $\mathcal{C}_4$ and $\mathcal{T}$ breaking mass term. In Ref. [\citen{huang2018higher}], Huang and Liu have proposed a binary drive while the Hamiltonian, in one step of the drive, accommodates corner modes in the static case. Vega {\it et.al.} in Ref. [\citen{rodriguez2019higher}] have studied a driven model with mirror symmetries that can harbor Floquet topological quadrupole phases and they have used a two step drive protocol to obtain higher order topological insulating phases. On the other hand, Ref. [\citen{PhysRevB.99.045441}] has reported a theoretical proposal for constructing static and Floquet SOTI by stacking 1D topological phases and coupling them with dimerized hopping amplitude (coupled wire construction). Therefore it is natural to ask whether such higher order phase hosting static as well as Floquet corner modes can also be obtained by periodically driving a semimetallic phase based on a model system where the drive is an external irradiation with circular polarized nature. It can also be an interesting fact to investigate whether one can realize first order TI phase as an intermediate phase before turning up to a HOTI phase. Moreover, the nature and stability of such phases with the enhancement of the driving strength of the external irradiation is worth to explore.

Motivated by the above-mentioned questions, in this work, we consider a two dimensional semimetal in presence of external irradiation. We show that the semimetal becomes Floquet topological insulator (FTI), characterized by quantized non-zero Chern number, under the influence of external irradiation for \eg circularly polarized light which breaks time-reversal symmetry $\mathcal{T}$. We derive the quasi-static Floquet Hamiltonian employing Brillouin-Wigner (BW) perturbation theory~\cite{mikami2016brillouin,arijitSilicene}. The sharp transition of the Chern number~\cite{fukui2005chern} from $+1$ to $-1$ takes place with a concomitant band gap closing in the bulk of the quasi energy spectrum. In the non-irradiated case, breaking the crystalline $\mathcal{C}_4$ symmetry by adding a mass perturbation, we show that the system becomes a static SOTI having in-gap corner modes while the bulk and the edges remain gapped. This static SOTI phase is identified by vanishing dipole moment and a half-integer quadrupole moment $Q_{xy}^{(0)}=0.5$ (modulo 1). As a result of the periodic drive on the SOTI phase, one can realize a Floquet second order topological insulator (FSOTI) phase where the Floquet corner modes are found to be pristine at all four corners of our system. Numerically computed Floquet quadrupole moment $Q_{xy}$ also turns out to be half-integer value (modulo 1) for the driven case. We also show the emergence of the dressed corner modes at finite quasi-energy $\omega/2$ (origin shifted by $\omega/2$), where $\omega$ is the driving frequency. These modes manifest the quasi-equilibrium nature of the FHOTI phase.

The remainder of the paper is organized as follows. In Sec.~\ref{sec:II}, we describe the model Hamiltonian for our setup, the driving protocol and formalism and present a brief outline of the derivation of the effective Hamiltonian in the high frequency regime. The results are presented in Sec.~\ref{sec:III} where we discuss the static semimetal Hamiltonian, the first order FTI and then both static SOTI and FSOTI phases in high frequency as well as in the intermediate frequency regime. Finally, we summarize our results and conclude in Sec.~\ref{sec:IV}.

%------------------------------------------------
\section{Model and method \label{sec:II}}
%-------------------------------------------------
\subsection {Static Hamiltonian}
%------------------------------------------
We begin with a 2D square lattice which describes a semimetal, for which the Hamiltonian reads

\begin{equation}\label{HSM}
H_{\rm SM}=\sum_{j,k}^{}\Big[c^\dagger_{j,k} T_x c_{j+1,k}+c^\dagger_{j,k} T_y c_{j,k+1} +\textrm{h.c.} \Big]\ ,
\end{equation}
with
\begin{eqnarray}
T_x&=&\frac{i t_1}{2} \Gamma_1+\frac{ t_2}{2} \Gamma_3 \ , \non\\ 
T_y&=&\frac{i t_1}{2} \Gamma_2+\frac{ t_2}{2} \Gamma_3 \ .
\end{eqnarray}

Here, $t_{1}$, $t_{2}$ are the amplitudes of nearest-neighbour hopping, $c_{j,k}$ is a four component spinor $\{A_\ua,B_\ua,A_\da,B_\da\}^{T}$ where $A,\, B$ 
are the orbitals and $i$, $j$ run along $x$ and $y$-directions respectively. The mutually anti-commuting hermitian $\Gamma$ matrices are: $\Gamma_1=\sigma_3 \tau_1$, $\Gamma_2=\sigma_0 \tau_2$, 
$\Gamma_3=\sigma_0 \tau_3$. Two sets of Pauli matrices ${\boldsymbol \tau}$ and ${\boldsymbol \sigma}$ respectively indicate the orbital and spin degrees of freedom of the system. \\

%~~~~~~~~~~~~~~~~~~~~~~~~~~~~~~~~~~~~~~~~~~~~~~~~~~~~~~~~~
\subsection {Driving protocol and Formalism}
\begin{figure}[!thpb]
	\includegraphics[width=0.4\textwidth]{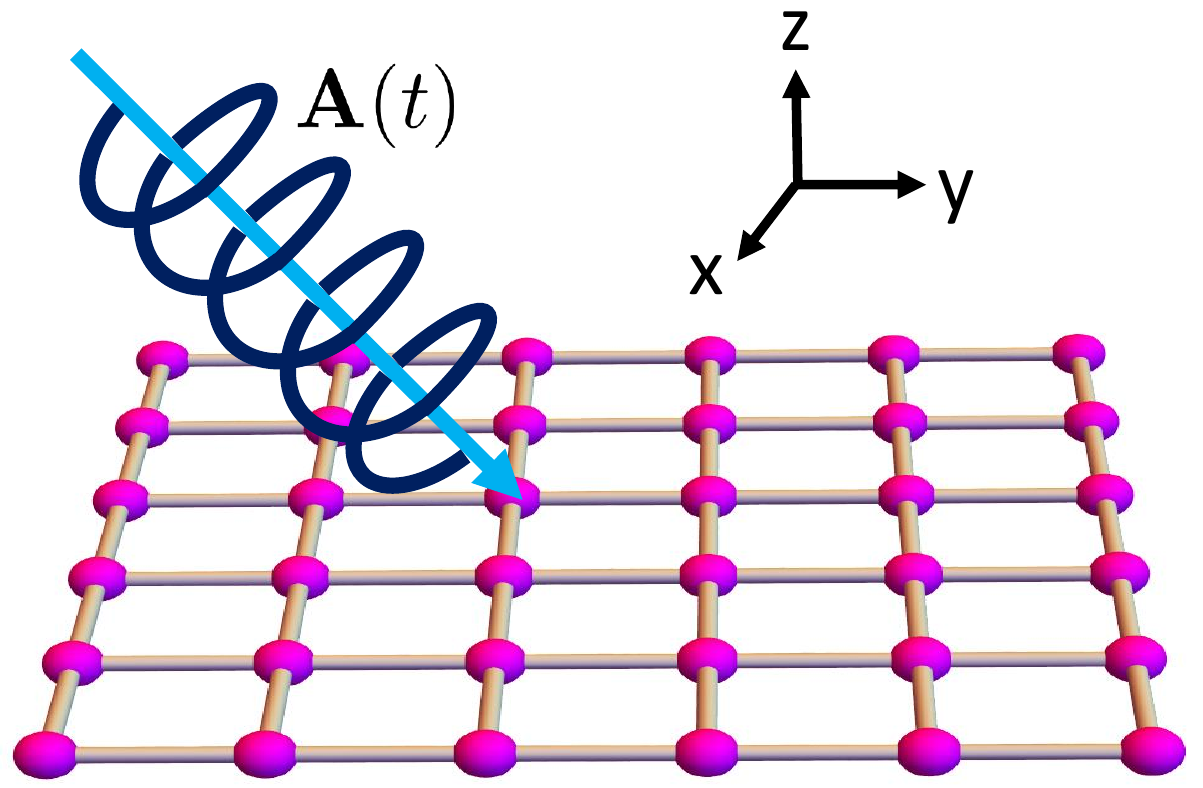}
	\caption{(Color online)Schematics of our square lattice geometry is shown in the presence of external irradiation. Here, the magenta (light grey) solid circles represent a square lattice and the blue (dark grey) helix denotes the external circularly polarized light modeled as a vector potential $A(t)$. 
	}
	\label{model}
\end{figure}

%~~~~~~~~~~~~~~~~~~~~~~~~~~~~~~~~~~~~~~~~~~~~~~~~~~~~~~~~~
The schematic of our geometry in the presence of external irradiation is demonstrated in Fig.~\ref{model}. We consider circularly polarized light of the form $\mathbf{A} (t)=\mathbf{A} (\cos(\omega t),\sin(\omega t))$ and the beam spot is much larger than the system in order to get rid of any spatial dependency. The purpose of choosing circularly polarised light instead of linearly polarised light is to break the time reversal symmetry and enabling us to explore non-trivial topological phases~\cite{arijitSilicene,usaj2014irradiated,perez2014floquet}. We begin with a time periodic Hamiltonian $H(t + T) = H(t)$ given by its Fourier components as
\begin{align}
\mathcal{H}_{n} = \int_{0}^{T}\frac{dt}{T} H(t) e^{in \omega t}\ , \label{hf}
\end{align}
where $T=2\pi/\omega$ is the period of the drive. In the frequency domain, the time independent infinite dimensional Hamiltonian can be written in the extended Floquet basis as~\cite{eckardt2015high}
\begin{widetext}
	\begin{equation}\label{FullFloquet}
	H_F=\begin{pmatrix}.\\
	
	& \mathcal{H}_0 -2\omega & \mathcal{H}_1 & \mathcal{H}_2 & .\\
	&  \mathcal{H}_{-1} & \mathcal{H}_0 -\omega & \mathcal{H}_1 & \mathcal{H}_2&.\\
	& \mathcal{H}_{-2} & \mathcal{H}_{-1} & \mathcal{H}_0 & \mathcal{H}_1 & \mathcal{H}_2&.\\
	& .&\mathcal{H}_{-2} & \mathcal{H}_{-1} & \mathcal{H}_0 +\omega & \mathcal{H}_1 & \mathcal{H}_2&.\\
	& & . &  \mathcal{H}_{-2} & \mathcal{H}_{-1} & \mathcal{H}_0 +2\omega & \mathcal{H}_1 &\mathcal{H}_2& .\\
	& & & & & & & &.\\	
	\end{pmatrix}
	\end{equation}
\end{widetext}
where, $\mathcal{H}_n$'s are defined in Eq. (\ref{hf}). The effect of the periodic drive is taken into account by considering a Peierls phase substitution in the hopping elements as
\begin{eqnarray}
T_x &\rightarrow& T_x e^{-i  A \cos(\omega t)}\ , \non \\
T_y &\rightarrow& T_y e^{-i  A \sin(\omega t)}\ .
\end{eqnarray} 

%~~~~~~~~~~~~~~~~~~~~~~~~~~~~~~~~~~~~~~~~~~~~~~~~~~~~~~~~~~~~~~~~~~~~~~~~~~
\subsection {Effective Hamiltonian in high frequency limit : Brillouin-Wigner perturbation expansion}
%~~~~~~~~~~~~~~~~~~~~~~~~~~~~~~~~~~~~~~~~~~~~~~~~~~~~~~~~~~~~~~~~~~~~~~~~~~
To begin with, we use Brillouin-Wigner (BW) perturbation theory~\cite{mikami2016brillouin} to obtain the effective Hamiltonian for the periodically driven system in the high-frequency limit \ie frequency is large compared to the bandwidth ($\omega\gg t$). Following Mikami \etal~\cite{mikami2016brillouin}, the effective Hamiltonian can be obtained in powers of $1/\omega$ using the BW perturbation theory. 
Although we consider here, terms only up to the order of $1/\omega$ for simplicity, it is important to note that the essential physics can be extracted curtailing the Hamiltonian up to $1/\omega$ in the high frequency limit as higher order terms in the $1/\omega$ expansion are vanishingly small. The effective Hamiltonian can be written as

\begin{equation}
\mathcal{H}_{\textrm{BW}}=\sum_{r=0}^{\infty} \mathcal{H}_{\textrm{BW}}^{(r)}\ ,
\label{H}
\end{equation}
with
\begin{eqnarray}
\mathcal{H}_{\textrm{BW}}^{(0)}&=&\mathcal{H}_0\ , \non\\
\mathcal{H}_{\textrm{BW}}^{(1)} &=& \sum_{n \neq 0} \frac{\mathcal{H}_{-n}\mathcal{H}_n}{n\omega}\ , \non\\
\mathcal{H}_{\textrm{BW}}^{(2)} &=& \mathcal{O}\Big(\frac{1}{\omega^2}\Big)\ .
\end{eqnarray}
The zeroth order Hamiltonian $\mathcal{H}_{0}$ contains terms as the unperturbed one, but with modulated hopping amplitudes whereas, the term with $O(1/\omega)$ manifests hopping elements originating due to the effect of irradiation in the high frequency limit~\cite{mikami2016brillouin, arijitSilicene,usaj2014irradiated}. 

%~~~~~~~~~~~~~~~~~~~~~~~~~~~~~~~~~~~~~~~~~~~~~~~~~~~~~~~~~~~~~~~~
%~~~~~~~~~~~~~~~~~~~~~~~~~~~~~~~~~~~~~~~~~~~~~~~~~~~~~~~~~~~~~~~~
%~~~~~~~~~~~~~~~~~~~~~~~~~~~~~~~~~~~~~~~~~~~~~~~~~~~~~~~~~~~~~~~~
%~~~~~~~~~~~~~~~~~~~~~~~~~~~~~~~~~~~~~~~~~~~~~~~~~~~~~~~~~~~~~~~~

Using BW expansion, we compute the effective Floquet Hamiltonian for the semimetal in presence of the drive which is given by Eq.(\ref{H}), where

\begin{widetext}
	\begin{eqnarray}
	\mathcal{H}_{\textrm{BW}}^{(0)} &=&\sum_{j,k}^{}\Big[c^\dagger_{j,k} T_1 c_{j+1,k}+c^\dagger_{j,k} T_2 c_{j,k+1} +\textrm{h.c.} \Big]\ , \non\\    
	\mathcal{H}_{\textrm{BW}}^{(1)}&=&\sum_{j,k}^{}c^\dagger_{j,k} M c_{j,k}+\sum_{j,k}^{}\Big[c^\dagger_{j,k} T_3 c_{j+1,k+1} +c^\dagger_{j,k} T_4 c_{j+2,k} +c^\dagger_{j,k} T_5 c_{j,k+2} +c^\dagger_{j,k} T_6 c_{j-1,k+1} +\textrm{h.c.} \Big]\ ,
	\label{FHam}    
	\end{eqnarray}
	with
	\begin{eqnarray}
	M&=&J_2\big(t_1^2+t_2^2\big) \sigma_0\tau_0\ , \non\\
	T_1&=& \frac{\mathcal{J}_0(A)}{2}\big(i t_1 \sigma_3\tau_1 +  t_2 \sigma_0 \tau_3 \big)\ , \non \\
	T_2&=& \frac{\mathcal{J}_0(A)}{2}\big(i t_1 \sigma_0\tau_2  + t_2 \sigma_0 \tau_3\big)\ , \non \\
	T_3&=&  J_{c1} t_2^2 \sigma_0\tau_0+J_{s1}\big(t_1^{2}\sigma_3\tau_3+it_1t_2\sigma_3\tau_2+it_1t_2\sigma_0\tau_1\big)\ ,  \non \\
	T_4&=& J_1\big(t_2^2-t_1^2 \big) \sigma_0\tau_0\ , \non\\
	T_5&=& J_1\big(t_2^2-t_1^2 \big)\sigma_0\tau_0\ ,     \non \\
	T_6&=&  J_{c2} t_2^2 \sigma_0\tau_0-J_{s2}\big(t_1^{2}\sigma_3\tau_3+it_1t_2\sigma_3\tau_2-it_1t_2\sigma_0\tau_1\big)\ ,  
	\label{hpr}
	\end{eqnarray}
	%\vspace{0.2 in}
	where,
	\begin{eqnarray}
	J_1&=&\sum_{n \neq0}\frac{(-1)^n\mathcal{J}_n^2(A)}{4n\omega},\quad \quad J_{c1}=\sum_{n \neq 0} \frac{(-1)^n \mathcal{J}_n^2(A) \cos\big(\frac{n \pi}{2}\big) }{2n\omega},\quad \quad J_{s1}=\sum_{n \neq 0} \frac{(-1)^n \mathcal{J}_n^2(A) \sin\big(\frac{n \pi}{2}\big) }{2n\omega}\non\\
	J_2&=&\sum_{n \neq 0}\frac{\mathcal{J}_n^2(A)}{n\omega},\quad \quad \quad \quad \, J_{c2}=\sum_{n \neq 0} \frac{\mathcal{J}_n^2(A) \cos\big(\frac{n \pi}{2}\big) }{2n\omega},\quad \quad \quad \quad \,\, J_{s2}=\sum_{n \neq 0} \frac{ \mathcal{J}_n^2(A)\sin\big(\frac{n \pi}{2}\big) }{2n\omega}.\non
	\end{eqnarray}
	Here, $\mathcal{J}_n$ is the Bessel function of first kind and $A$ is the amplitude of the drive.
\end{widetext} 

\begin{figure}[!thpb]
	\includegraphics[width=0.3\textwidth]{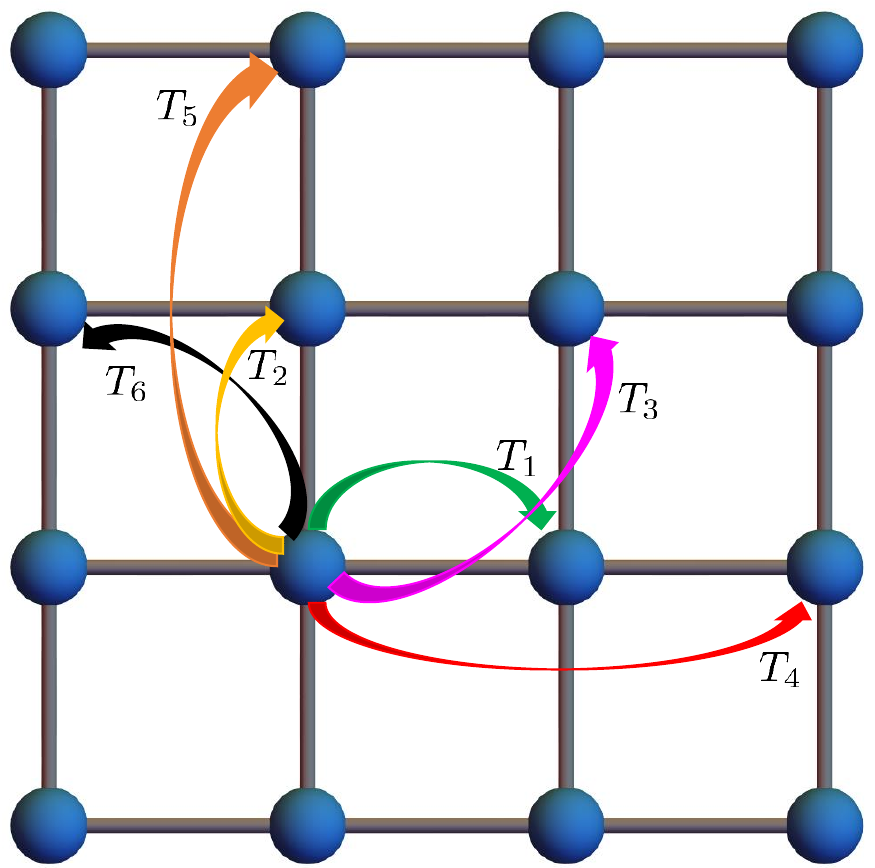}
	\caption{(Color online) Directions of newly generated effective hoppings are depicted by different arrows in the square lattice system. Here, $T_1$ and $T_2$ represent the renormalized 
		nearest-neighbour hopping elements, whereas $T_3$, $T_4$, $T_5$ and $T_6$ denote the drive induced next nearest neighbour hoppings in different directions within the square lattice. 
	}
	\label{Hopping}
\end{figure}

From Eq.(\ref{HSM}), it is evident that only nearest neighbor hoppings are present in the static Hamiltonian describing a semimetal. Here $T_1$ and $T_2$ are the renormalized amplitudes of such hoppings. On the other hand, $T_3$, $T_4$, $T_5$ and $T_6$ are the newly generated next nearest neighbour hoppings in different directions. Such long-range hopping generation by the periodic drive is found earlier as well in the other systems~\cite{arijitSilicene,PhysRevB.87.201109}. We have shown all these hoppings schematically in Fig.~\ref{Hopping} for our system.

It is worth to mention here that the irradiation effect in the high frequency limit can also be taken into account by using other perturbation expansion methods \ie\, Floquet Magnus~\cite{casas2001floquet,blanes2009magnus}, van Vleck perturbation theory~\cite{eckardt2015high}. Floquet Magnus expansion harbors unwanted driving phase dependence as a result of which the effective Hamiltonian also contains the driving phase. On the other hand, it is much more complicated to write higher order terms in van Vleck perturbative expansion series. 

%~~~~~~~~~~~~~~~~~~~~~~~~~~~~~~~~~~~~~~~~~~~~~~~~~~~~~~~~~~~~~~
%~~~~~~~~~~~~~~~~~~~~~~~~~~~~~~~~~~~~~~~~~~~~~~~~~~~~~~~~~~~~~~
%~~~~~~~~~~~~~~~~~~~~~~~~~~~~~~~~~~~~~~~~~~~~~~~~~~~~~~~~~~~~~~
%~~~~~~~~~~~~~~~~~~~~~~~~~~~~~~~~~~~~~~~~~~~~~~~~~~~~~~~~~~~~~~

%-------------------------------------------------------------------------------------
\section{Results}{\label{sec:III}}
%-------------------------------------------------------------------------------------
In this section, we present our results involving static phases as well as Floquet phases. Starting from the static semimetal phase, we discuss the FTI and then analyze both static and Floquet SOTI. 
At first we discuss the high frequency limit and finally, we show the appearence of dressed corner modes in the Floquet SOTI phase considering intermediate frequency regime. 
We have considered both $t_1, t_2=1$ for all our numerical results and all the parameters having the dimension of energy are scaled with respect to the hopping strength. For all the results shown below,
in the high frequency limit, we have chosen $\omega=10$.

%-----------------------------------------------------
\subsection {Trivial static semimetal without drive}
\begin{figure}[H]
	\centering{
		\includegraphics[width=0.23\textwidth, height=3.8cm, clip=true]{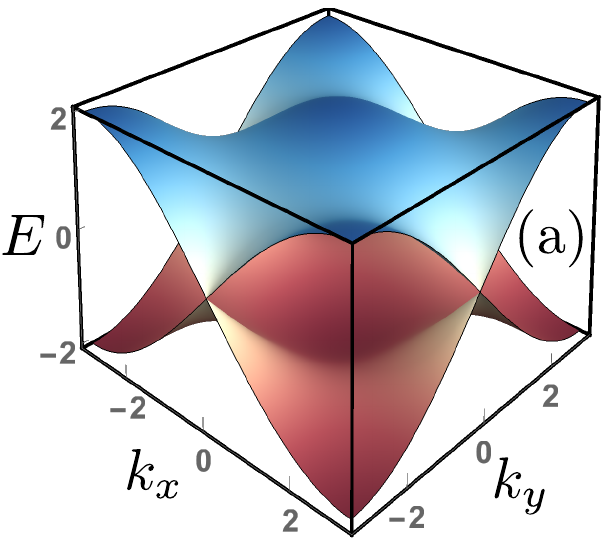}\hfill
		\includegraphics[width=0.23\textwidth, height=3.8cm, clip=true]{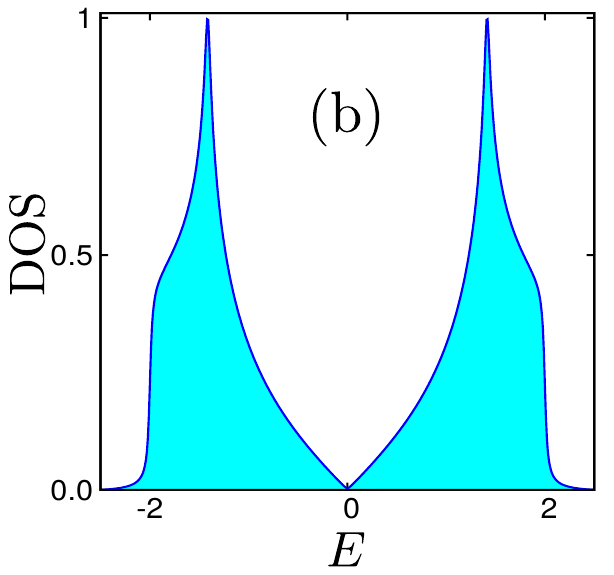}}
	\caption{(Color online) (a) Bulk band spectrum for the static semimetal phase is shown. (b) Total DOS of the semimetal phase is demonstrated.
	}
	\label{Band}
\end{figure}

%-----------------------------------------------------
The bulk band structure of the static Hamiltonian given by Eq.(\ref{HSM}) is shown in Fig.~\ref{Band}(a) and the corresponding total density of states (DOS) is presented in Fig.~\ref{Band}(b). As we see from Fig.~\ref{Band}(a), the conduction and valence band meet at four points in the first Brillouin zone : $\Gamma =(0,0), X=(0,\pi), Y=(\pi,0) \textrm{ and } S=(\pi,\pi)$, describing a semimetallic behavior. 
The low energy spectrum at those four points exhibits a massless Dirac like dispersion. Therefore, this phase corresponds to a trivial semimetal with Chern number $\rm zero$. The DOS plot resembles similar to that of graphene~\cite{neto2009electronic} as it is vanishingly small near the zero of the Fermi energy (Dirac point in case of graphene) and scales linearly around it.
%----------------------------------------------------------------
%----------------------------------------------------------------
%----------------------------------------------------------------
%----------------------------------------------------------------

%-------------------------------------------------------
\subsection {First order topological insulator with drive}
%-------------------------------------------------------
We now demonstrate the outcome of the external irradiation on the semimetal Hamiltonian (Eq.(\ref{HSM})). As the circularly polarized light breaks time-reversal symmetry $\mathcal{T}$ in the semimetal phase, it becomes an FTI with chiral edge states in presence of the external irradiation. In the high frequency regime, the effective Floquet Hamiltonian in the presence of the external drive is given by Eq.(\ref{FHam}).

%~~~~~~~~~~~~~~~~~~~~~~~~
\subsubsection{Band topology}
%~~~~~~~~~~~~~~~~~~~~~~~~
In presence of the external drive, the band topology of the system exhibits very intriguing behavior. To understand this, we explore the bulk quasi-energy spectrum along the high symmetry line: $\Gamma-X-S-Y-\Gamma$ for three different values of driving amplitude $A$. They are illustrated in Fig.~\ref{G}[(a),(b),(c)]. A finite bandgap is present in bulk quasi-energy spectrum for $A=3.0$ as shown in Fig.~\ref{G}(a), while the bandgap closes at $A=3.35$ (see Fig.~\ref{G}(b)). The bandgap again starts to reopen which is depicted in Fig.~\ref{G}(c) for $A=3.8$. The value of $A$ at which the bulk bandgap closes, 
Chern number ($C_n$) changes sign (see text for discussion). The band inversion process, with the rise of the strength of the external drive, can be understood as a Floquet topological phase transition in the bulk quasi-energy spectrum.  Similar band inversion again takes place near $A=7.3$ (not shown here) where $C_n$ changes sign (see Fig~\ref{G}(d)).

\vspace{0.5cm}
%~~~~~~~~~~~~~~~~~~~~~~~~~~~~~~~~~~~~~~
\subsubsection{Chern number and edge modes}
%~~~~~~~~~~~~~~~~~~~~~~~~~~~~~~~~~~~~~~
%-------------------------------------------------------------------------
%-------------------------------------------------------------------------
\begin{figure*}[!thpb]
	\hspace*{\fill}%
	\includegraphics[width=1.0\textwidth, height=10.0cm, clip=true]{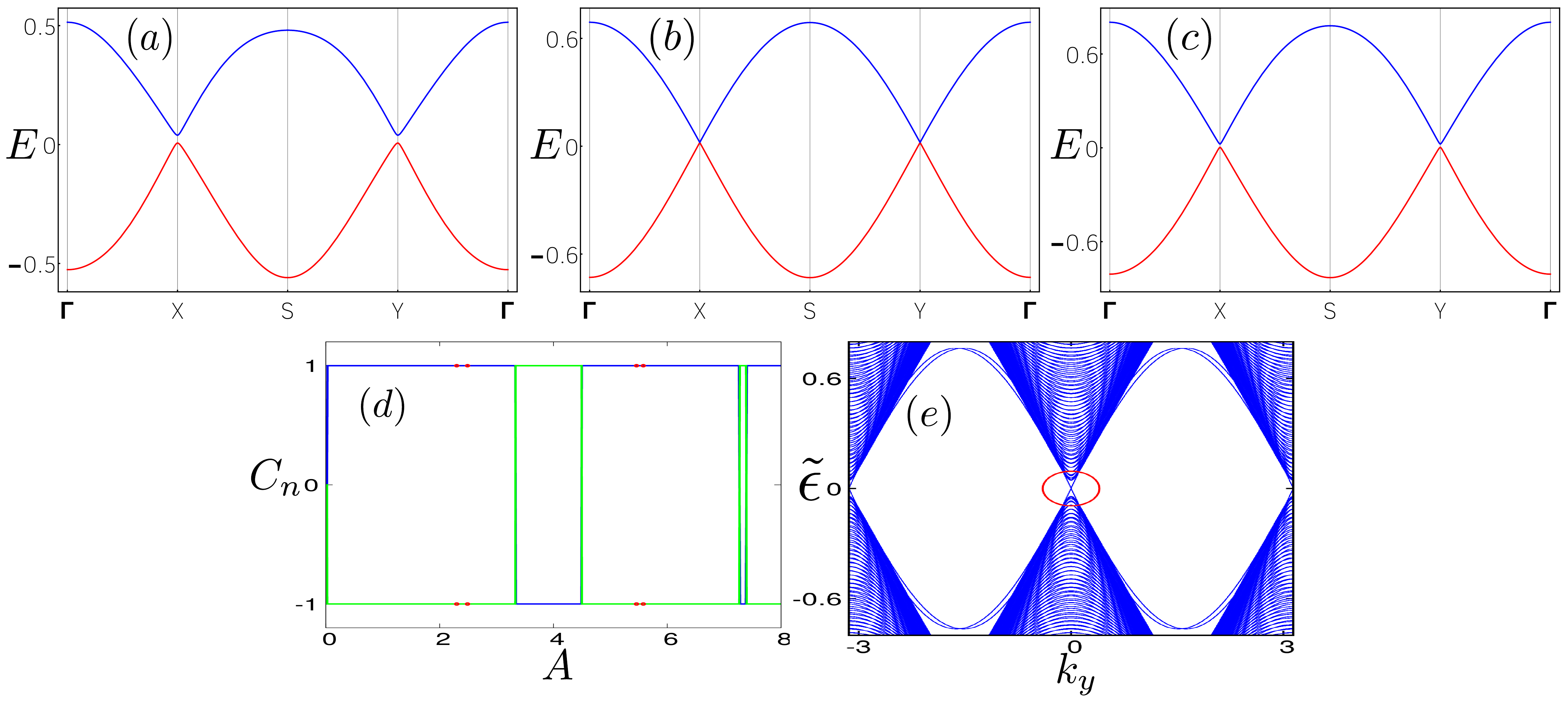}\hfill
	\hspace*{\fill}%
	\caption{(Color online)  Panel (a),(b),(c): Bulk band structure is shown along the $\Gamma-X-S-Y-\Gamma$ line, indicating the presence of a bulk bandgap, gap closing and reopening for $A =3,\,3.35,\,3.8$ respectively. (d): Chern number is shown as a function of driving the amplitude,  $A$. Unphysical points where the Hamiltonian vanishes are indicated by red dots. (e): Band structure for a $y$-directed slab with counter propagating dispersive Floquet edge modes is presented for $A =1.0$ and $N_x=100$. Floquet edge modes around $k_{y}=0$ are marked by the red circle.
	}
	\label{G}
\end{figure*}
%-------------------------------------------------------------------------
%--------------------------------------------------------------------------

To identify the Floquet topological phase transition, we calculate the Chern number $C_{n}$ of the bulk band following Ref.~[\onlinecite{fukui2005chern}]. The Chern number of the upper and lower band is shown as a function of driving amplitude $A$ in Fig.~\ref{G}(d). As the periodic drive is turned on ($A\neq 0$), due to broken $\mathcal{T}$, we see a sudden jump of $C_n$ from $0$ to $1$ for the upper band and $0$ to $-1$ for the lower band where the system becomes an FTI with counter propagating dispersive edge modes. We present the band structure in a $y$-directed slab geometry in Fig.~\ref{G}(e) where the edge modes are visible for $A=1.0$. Similar edge modes have been found in the case of graphene which becomes an FTI in presence of external periodic drive~\cite{usaj2014irradiated,perez2014floquet}.

The first jump of $C_n$ for the upper band (lower band) from $+1$ to $-1$ ($-1$ to $+1$) takes place for $A=3.35$ where the band inversion occurs. Similar Floquet topological phase transition also occurs 
for higher values of $A$ ($A=7.3$), as can be seen from Fig.~\ref{G}(d), which again brings the concomitant band inversion. Note that, the value of $C_n$ is ill-defined near the points $A=2.4$ and $5.5$ where the Hamiltonian itself is vanishingly small due to the vanishing of Bessel function $\mathcal{J}_0$~\cite{mikami2016brillouin}. Those points are marked by red dots in Fig.~\ref{G}(d). 
This behavior persists even one takes into consideration the higher order terms in $1/\omega$~\cite{mikami2016brillouin}. 

%------------------------------------------------------------------------
\subsection {Higher order topological phase with high frequency drive}
%------------------------------------------------------------------------
We now move to the main part of the paper where we describe the appearance of the HOTI phase. We add a Wilson mass term of the form~\cite{schindler2018higher,PhysRevB.100.115403,agarwala2019higher,nag2019higher} $H_B=\Delta [\cos k_x -\cos k_y ] \sigma_1 \tau_1$ with the initial Hamiltonian $H_\textrm{SM}$ given by Eq.(\ref{HSM}). $H_B$ breaks both $\mathcal{C}_4$ rotational symmetry and $\mathcal{T}$ but preserves the combined symmetry operation $\mathcal{C}_4 \mathcal{T}$. We note that $\{H_{SM},H_B\}=0$. ${H_B}$ changes sign under $\mathcal{C}_4$ rotation. Effective 1D Dirac equation for the edge(s) of the system contains a mass term that changes sign across the corner~\cite{yan2018majorana,wang2018high} and thus the presence of four corner modes is ensured by a generalized Jackiw-Rebbi index theorem~\cite{jackiw1976solitons}. A static second order topological insulator accommodating in-gap corner modes at all the four corners can thus be realized in the square lattice system. Similar mass term has been used recently to obtain higher order topological phases in various models~\cite{schindler2018higher,agarwala2019higher,nag2019higher}.

In this subsection, we present our results for circularly polarized light with high frequency. In presence of the irradiation, the effective Floquet Hamiltonian is given by Eq.(\ref{H}) and Eq.(\ref{FHam}) including the new terms arising due to the $\mathcal{C}_4$ symmetry breaking mass perturbation. Different hopping amplitudes in Eq.(\ref{hpr}) are now modified by the mass parameter $\Delta$ and given by,

\begin{widetext}
	\begin{eqnarray}
	M&=&J_2\big(t_1^2+t_2^2+\Delta^2 \big) \sigma_0\tau_0\ , \non\\
	T_1&=& \frac{\mathcal{J}_0(A)}{2}\big(i t_1 \sigma_3\tau_1  + t_2 \sigma_0 \tau_3+\Delta \sigma_1 \tau_1 \big)\ , \non \\
	T_2&=& \frac{\mathcal{J}_0(A)}{2}\big(i t_1 \sigma_0\tau_2  + t_2 \sigma_0 \tau_3-\Delta \sigma_1 \tau_1\big)\ , \non \\
	T_3&=&  J_{c1}\big(t_2^2-\Delta^2\big)\sigma_0\tau_0+J_{s1}\big(t_1^{2}\sigma_3\tau_3+it_1t_2\sigma_3\tau_2+it_1t_2\sigma_0\tau_1+2t_2\Delta\sigma_1\tau_2+it_1\Delta\sigma_2\tau_0-it_1\Delta\sigma_1\tau_3\big)\ ,  \non \\
	T_4&=& J_1\big(t_2^2-t_1^2+\Delta^2 \big) \sigma_0\tau_0\ , \non\\
	T_5&=& J_1\big(t_2^2-t_1^2+\Delta^2 \big)\sigma_0\tau_0\ ,     \non \\
	T_6&=&  J_{c2}\big(t_2^2-\Delta^2\big)\sigma_0\tau_0-J_{s2}\big(t_1^{2}\sigma_3\tau_3+it_1t_2\sigma_3\tau_2-it_1t_2\sigma_0\tau_1-2t_2\Delta\sigma_1\tau_2+it_1\Delta\sigma_2\tau_0+it_1\Delta\sigma_1\tau_3\big)\ .  
	\label{hotih}
	\end{eqnarray}
\end{widetext}
where different parameters have their usual meaning as defined earlier. 

%~~~~~~~~~~~~~~~~~~~~~~~~~~~
\subsubsection{Quadrupole moment}
%~~~~~~~~~~~~~~~~~~~~~~~~~~~
To establish the HOTI phase in our model we first investigate the topological invariant. From the existing literature~\cite{benalcazar2017quantized,benalcazar2017electric}, it is established that SOTIs are distinguished by vanishing dipole moment but exhibiting quantized quadrupole moment $Q_{xy}=0.5$ (modulo 1)~\cite{wheeler2018many,kang2018many,agarwala2019higher,nag2019higher}. 
Here we present the brief outline on how to calculate quadropolar moment numerically for the HOTI phase. The macroscopic quadrupole moment for a crystal obeying periodic boundary condition is defined as 
following~\cite{wheeler2018many,kang2018many} :
\begin{equation}\label{macroquad}
Q_{xy}^{(0)}=\frac{1}{2 \pi} \textrm{Im}[ \ln \bra{\Psi_0} e^{2 \pi i \sum_{r}\hat{q}_{xy}(\mathbf{r})}\ket{\Psi_0}]\ ,
\end{equation}
Here, $\hat{q}_{xy}(\mathbf{r})=\frac{xy}{L^2}\hat{n}(\mathbf{r})$ is the microscopic quadrupole moment at a lattice site $\mathbf{r}$ with respect to $x=y=0$, while $L$ is the lattice size and $\ket{\Psi_0}$ 
is a many-body ground state which can be defined using the occupied states through the Slater determinant~\cite{restaprl}.

%-------------------------------------------------------------------------
%-------------------------------------------------------------------------
\begin{figure*}[!thpb]
	\hspace*{\fill}%
	\includegraphics[width=0.8\textwidth, height=6.5cm, clip=true]{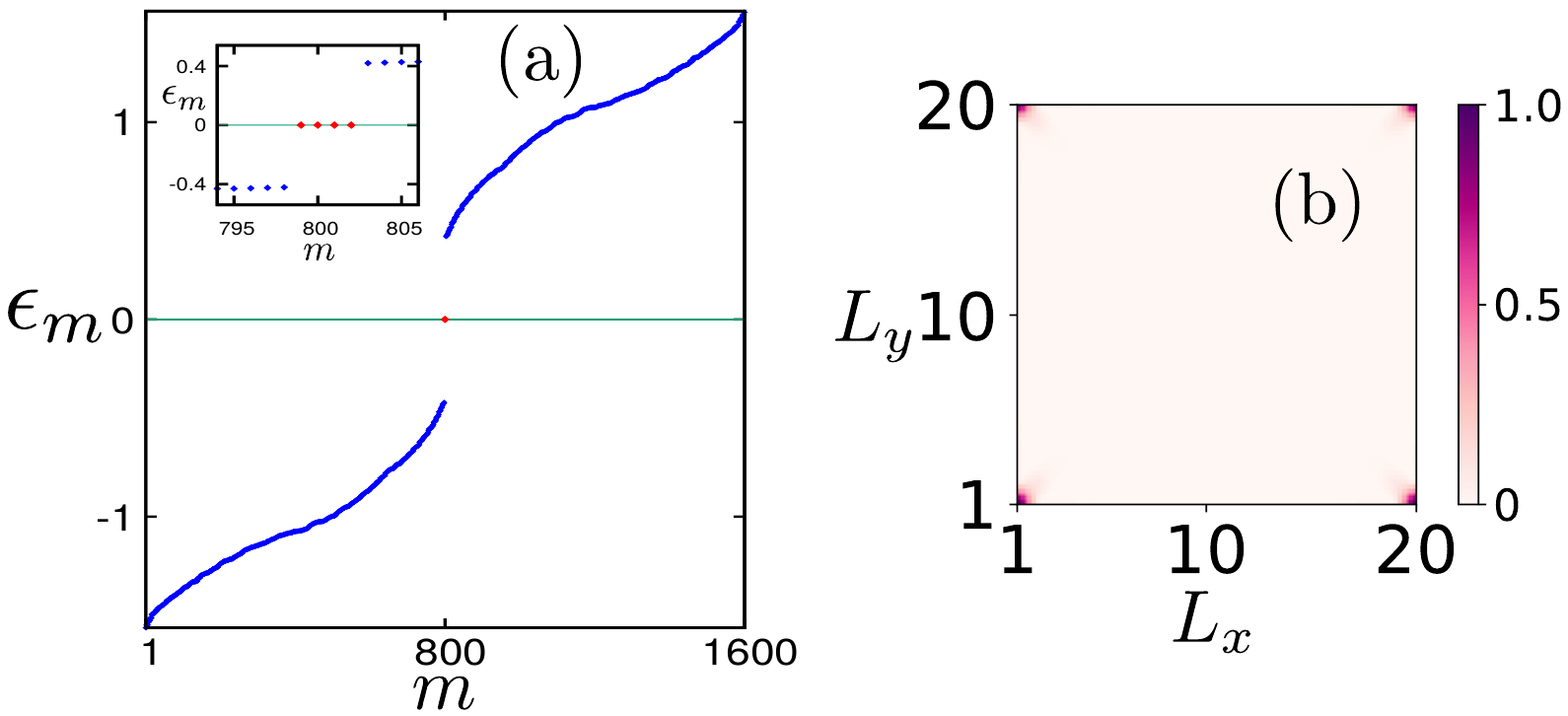}\hfill
	\hspace*{\fill}%
	\caption{(Color online) (a) The eigenvalue spectrum is shown as a function of the number of eigenstates where the red point indicates the zero energy states while the blue line denotes the quasi-energy bulk. Four in-gap zero energy states (corner modes) are illustrated by four red dots in the inset. (b) The four corner modes are manifested via LDOS for $A\neq 0$ at $E=0$.
	}
	\label{Mode}
\end{figure*}
%-------------------------------------------------------------------------
%--------------------------------------------------------------------------

To evaluate the quadrupole moment using Eq.(\ref{macroquad}), we construct a $N\times N_o$ dimensional matrix $\mathcal{U}$ by a column-wise marshaling of $N_o$ eigenvectors according to their energy. Here $N$ is the total number of states while $N_o$ is the number of occupied states. Afterward, we formulate another matrix operator $\mathcal{W}$ :  
\begin{align}\label{WforQuad}
\mathcal{W}_{ i\alpha,j} = \exp \left[ {i \frac{2\pi}{L^2}f(x_{i\alpha},y_{i\alpha})} \right]  \: \mathcal{U}_{ i\alpha,j }\ ,
\end{align}
Here, $\alpha$ includes both the orbial and spin index. We consider $f(x_{i\alpha},y_{i\alpha})=x_{i\alpha}y_{i\alpha}$. Therefore, $Q_{xy}^{(0)}$, defined in Eq.(\ref{macroquad}), can be recasted into the 
following form using the above mentioned matrices as  
\begin{align}
Q_{xy}^{(0)}=\frac{1}{2\pi} \;\textrm{Im} \;[\textrm{Tr}  \ln \left(\mathcal{U}^\dagger \mathcal{W} \right)]\ .
\end{align}

We obtain the value of quadrupole moment only up to modulo 1 and now on we omit the term ``modulo 1" but whenever we present the value of the quadrupole moment, we intend it as modulo 1. 
We compute $Q_{xy}^{(0)}$ when $\mathcal{C}_4$ is broken along with $\mathcal{T}$ where the concomitant topological static corner modes arise in the system. We find that $Q_{xy}^{(0)}$ is always quantized with the value of $0.5$ (within numerical accuracy) for our system is in second order topological insulating phase. 

With the inclusion of the external periodic drive, we now compute the Floquet quadropole moment which is given by~\cite{wheeler2018many,kang2018many,agarwala2019higher,nag2019higher} 
$Q_{xy} = Q_{xy}^{(0)}-Q_{xy}^{\rm a}$; where $Q_{xy}^{\rm a}= \frac{1}{2}\sum_{i\alpha} \frac{1}{L^2} f(x_{i\alpha},y_{i\alpha})$ is the value of $Q_{xy}$ in the atomic limit. To numerically evaluate the quadropole moment $Q_{xy}$ from the Floquet Hamiltonian given by Eq.(\ref{FHam}) along with the new hopping parameters (see Eq.(\ref{hotih})), we use the same prescription mentioned above. 
We find that $Q_{xy}$ is always quantized having a value of $0.5$ for any nonzero value of $A$ except when $\mathcal{J}_0 (A)$ becomes zero which in turn makes the Hamiltonian vanishingly small and thus making the numerical evaluation of $Q_{xy}$ ill-defined. 

Note that, as the drive is turned on, an onsite potential term is generated which is given by $M$ in the effective Floquet Hamiltonian in Eq.(\ref{FHam}). This term indicates the shifting of the band spectra in the quasi-energy space. The effect of the drive generated onsite potential can always be nullified by implementing an extra suitable gate voltage connected to the 2D system. We have neglected this term in our calculation for simplicity and a better understanding of the engineering of the Floquet HOTI phase. The latter has also been reported very recently via the application of quench~\cite{nag2019higher}. 

%~~~~~~~~~~~~~~~~~~~~~~~~~~~~
\subsubsection{Floquet corner modes}
%~~~~~~~~~~~~~~~~~~~~~~~~~~~~

The static SOTI phase hosts in gap corner modes~\cite{benalcazar2017quantized,benalcazar2017electric}. Similarly, the FSOTI is characterized by the appearance of Floquet corner modes~\cite{huang2018higher,rodriguez2019higher,PhysRevB.99.045441}. The signature of these zero energy corner modes appears in the local density of states (LDOS). The LDOS as a function of energy $E$ at $i$-th site of a lattice is defined as
\begin{equation}\label{LDOS}
\rho_i(E)=\sum_{\lambda}^{}\lvert \braket{i|n_\lambda}\rvert^2\delta(E-\lambda) \ , 
\end{equation}
where, $\ket{i}$ and $\ket{n_\lambda}$ are the eigenstates of the Hamiltonian. To probe the zero energy corner the modes we have calculated LDOS at $E=0$. The Floquet corner modes that arise 
in our square lattice system with the periodic drive ($A \neq 0$) and $C_{4}$ symmetry breaking term, are shown in Fig.~\ref{Mode}(b), where the LDOS is depicted along two spatial directions 
($L_{x}$, $L_{y}$) of the sample. One can observe that the corner modes are almost (due to finite system size) localized at the four corners of the system. In Fig.~\ref{Mode}(a), we have shown the eigenvalue spectrum of the Floquet Hamiltonian. The in-gap corner modes (at zero energy) are shown by the red dot while the blue line indicates the continuum bulk. Four corner modes, all of which are at zero energy, are depicted in the inset of Fig.~\ref{Mode}(a). We emphasize the fact that the 0D Floquet corner modes are robust against the high frequency drive and are pinned at zero energy. In case of Floquet corner mode, we have calculated the quadrupole moment, $Q_{xy}$ using the eigenstates of the Hamiltonian in Eq.(\ref{hotih}) and $Q_{xy}$ always remain at $0.5$ irrespective of the driving strength $A$. 
Although in this frequency domain, one cannot distinguish between the static and Floquet corner modes, while the later arises due to the virtual photon transitions between the highest 
Floquet sub-bands.

%~~~~~~~~~~~~~~~~~~~~~~~~~~~~~~~~~~~~~~~~~~~~~~~~~~~~~~~~~~~~~~~~~~
\subsection {Higher order topological phase with intermediate frequency drive}
%~~~~~~~~~~~~~~~~~~~~~~~~~~~~~~~~~~~~~~~~~~~~~~~~~~~~~~~~~~~~~~~~~~
To incorporate the real photon transitions between different Floquet sub-bands, in this subsection, we consider the intermediate frequency regime and obtain the corner modes termed as 
{\it {dressed corner modes.}} As it is a formidable task to diagonalise an infinite dimensional Floquet Hamiltonian given in Eq.(\ref{FullFloquet}), we truncate the Hamiltonian upto four
Floquet-zone sector \ie we restrict ourselves to $m=0,1,2,3$ subspace~\cite{peng2019higher,PhysRevLett.123.016806}. This truncation of the infinite dimensional Hamiltonian is familiar 
in the context of intermediate frequency~\cite{usaj2014irradiated,perez2014floquet,PhysRevLett.123.016806}. The reduced Floquet Hamiltonian reads

\begin{widetext}
\begin{equation}\label{midHam}
\tilde{H}_F=\begin{pmatrix}
\mathcal{H}_0-\frac{7\omega}{2} & \mathcal{H}_1 & \mathcal{H}_2 & \mathcal{H}_3 & 0 & 0 & 0\\
\mathcal{H}_{-1} & \mathcal{H}_0-\frac{5\omega}{2} & \mathcal{H}_1 & \mathcal{H}_2 &\mathcal{H}_3 & 0 & 0\\
\mathcal{H}_{-2} & \mathcal{H}_{-1} & \mathcal{H}_0-\frac{3\omega}{2} & \mathcal{H}_1 & \mathcal{H}_2 &\mathcal{H}_3 & 0 \\
\mathcal{H}_{-3} & \mathcal{H}_{-2} & \mathcal{H}_{-1} & \mathcal{H}_0-\frac{\omega}{2} & \mathcal{H}_1 & \mathcal{H}_2 &\mathcal{H}_3 \\
0&\mathcal{H}_{-3} & \mathcal{H}_{-2} & \mathcal{H}_{-1} & \mathcal{H}_0+\frac{\omega}{2} & \mathcal{H}_1 & \mathcal{H}_2 \\
0&0&\mathcal{H}_{-3} & \mathcal{H}_{-2} & \mathcal{H}_{-1} & \mathcal{H}_0+\frac{3\omega}{2} & \mathcal{H}_1 \\
 0&0&0&\mathcal{H}_{-3} & \mathcal{H}_{-2} & \mathcal{H}_{-1} & \mathcal{H}_0+\frac{5\omega}{2}
\end{pmatrix}+\frac{\omega}{2} \mathbb{I}
\end{equation}
\end{widetext}

where,
\begin{eqnarray}
\mathcal{H}_0 &=&\sum_{j,k}^{}\Big[c^\dagger_{j,k} \mathcal{T}_1 c_{j+1,k}+c^\dagger_{j,k} \mathcal{T}_2 c_{j,k+1} +\textrm{h.c.} \Big]\ , \non\\    
\mathcal{H}_1 &=&\sum_{j,k}^{}\Big[c^\dagger_{j,k} \mathcal{T}_3 c_{j+1,k}+c^\dagger_{j,k} \mathcal{T}_4 c_{j,k+1} +\textrm{h.c.} \Big]\ , \non \\
\mathcal{H}_2 &=&\sum_{j,k}^{}\Big[c^\dagger_{j,k} \mathcal{T}_5 c_{j+1,k}+c^\dagger_{j,k} \mathcal{T}_6 c_{j,k+1} +\textrm{h.c.} \Big]\ , \non \\
\mathcal{H}_3 &=&\sum_{j,k}^{}\Big[c^\dagger_{j,k} \mathcal{T}_7 c_{j+1,k}+c^\dagger_{j,k} \mathcal{T}_8 c_{j,k+1} +\textrm{h.c.} \Big]\ ,
\label{FMidHam}    
\end{eqnarray}
with
\begin{eqnarray}
\mathcal{T}_1&=& \frac{\mathcal{J}_0(A)}{2}\big(i t_1 \sigma_3\tau_1  + t_2 \sigma_0 \tau_3+\Delta \sigma_1 \tau_1\big)\ , \non \\
\mathcal{T}_2&=& \frac{\mathcal{J}_0(A)}{2}\big(i t_1 \sigma_0\tau_2  + t_2 \sigma_0 \tau_3-\Delta \sigma_1 \tau_1\big)\ , \non \\
\mathcal{T}_3&=& \frac{\mathcal{J}_1(A)}{2}\big( t_1 \sigma_3\tau_1  -i t_2 \sigma_0 \tau_3-i\Delta \sigma_1 \tau_1\big)\ , \non \\
\mathcal{T}_4&=& \frac{\mathcal{J}_1(A)}{2}\big( i t_1 \sigma_0\tau_2  + t_2 \sigma_0 \tau_3-\Delta \sigma_1 \tau_1\big)\ , \non \\
\mathcal{T}_5&=& \frac{\mathcal{J}_2(A)}{2}\big(-i t_1 \sigma_3\tau_1  - t_2 \sigma_0 \tau_3-\Delta \sigma_1 \tau_1\big)\ , \non \\
\mathcal{T}_6&=& \frac{\mathcal{J}_2(A)}{2}\big(i t_1 \sigma_0\tau_2  + t_2 \sigma_0 \tau_3-\Delta \sigma_1 \tau_1\big)\ , \non \\
\mathcal{T}_7&=& \frac{\mathcal{J}_3(A)}{2}\big( -t_1 \sigma_3\tau_1  +i t_2 \sigma_0 \tau_3+i\Delta \sigma_1 \tau_1\big)\ , \non \\
\mathcal{T}_8&=& \frac{\mathcal{J}_3(A)}{2}\big( i t_1 \sigma_0\tau_2  + t_2 \sigma_0 \tau_3-\Delta \sigma_1 \tau_1\big)\ . 
\end{eqnarray}

Here, $\mathbb{I}$ is an identity matrix in the Floquet subspace basis. The last term in Eq.(\ref{midHam}) shifts the origin of energy by $\omega/2$ and have been neglected for our calculation~\cite{peng2019higher,PhysRevLett.123.016806}.

%{\textcolor{red}{We present our numerical results for this frequency regime in Fig.~\ref{Inter}. The outcome of intermediate frequency drive is the emergence of a gap opening at energy $\omega/2$~%%\cite{usaj2014irradiated,perez2014floquet,rudner2013anomalous,PhysRevLett.123.016806} \textcolor{green}{which cannot be captured by studying the high frequency periodic drive}. Such gap opening at %finite energy $\omega/2$ and the emergence of chiral edge states in the gap has been reported earlier for Graphene~\cite{usaj2014irradiated,perez2014floquet} and topological insulator~\cite{rudner2013anomalous}.}}

\begin{figure}[!thpb]
	\includegraphics[width=0.5\textwidth]{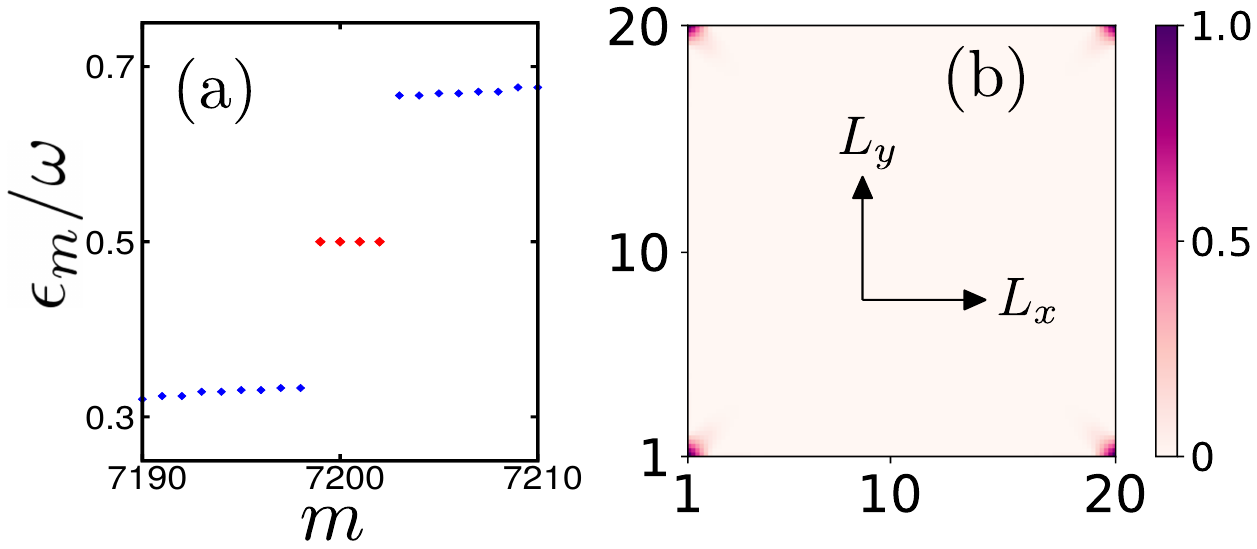}
	\caption{(Color online) (a) The quasienergy spectrum (where origin is shifted by $\omega/2$) is shown as a function of the number of eigenstates . (b) The four corner modes are manifested via LDOS 
for $A\neq 0$ at $\epsilon_m=\omega/2$. Here, we have chosen $A=0.5$ and $\omega=3.0$.
	}
	\label{Inter}
\end{figure}

We present our numerical results for this frequency regime in Fig.~\ref{Inter}. The quasienergy spectrum, around $\omega/2$, obtained from the exact diagonalization of the Hamiltonian, 
given in Eq.(\ref{midHam}), is presented in Fig.~\ref{Inter}(a) where one can note the presence of four zero energy modes exactly at the energy $\omega/2$. The signature of these dressed corner modes
at quasienergy $\omega/2$ are depicted in Fig.~\ref{Inter}(b) via  the LDOS. The latter is plotted along two spatial directions ($L_{x}$, $L_{y}$) of the system. Here, also the corner modes are almost 
localized at the four corners of the system. 
%~~~~~~~~~~~~~~~~~~~~~~~~~~~~~~~~~~~~~~~~~~~~~~~~~~~~~~~~~~~~~~~~
%~~~~~~~~~~~~~~~~~~~~~~~~~~~~~~~~~~~~~~~~~~~~~~~~~~~~~~~~~~~~~~~~
%~~~~~~~~~~~~~~~~~~~~~~~~~~~~~~~~~~~~~~~~~~~~~~~~~~~~~~~~~~~~~~~~
%~~~~~~~~~~~~~~~~~~~~~~~~~~~~~~~~~~~~~~~~~~~~~~~~~~~~~~~~~~~~~~~~
Note that, the dressed corner modes appearing at quasi-energy $\omega/2$ are actually the remnants of zero enrgy modes as we have neglected the 
last term in Eq.(\ref{midHam}) which shifts the origin of energy from 0 to $\omega$/2.

The quadrupole moment for the $m^{\rm th}$ Floquet zone can be defined as
\begin{equation}\label{floquetmidmacroquad}
Q_{xy,m}^{(0)}=\frac{1}{2 \pi} \textrm{Im}[ \ln \bra{\Psi_{0,m}} e^{2 \pi i \sum_{r}\hat{q}_{xy}(\mathbf{r})}\ket{\Psi_{0,m}}]\ .
\end{equation}
Here, we construct $\ket{\Psi_{0,m}}$ using the occupied states of $m^{\rm th}$ Floquet zone. 
Similarly, Eq.(\ref{WforQuad}) in this case becomes
\begin{align}\label{WforQuadmid}
\mathcal{W}_{ i\alpha,j,m} = \exp \left[ {i \frac{2\pi}{L^2}f(x_{i\alpha},y_{i\alpha})} \right]  \: \mathcal{U}_{ i\alpha,j,m }\ ,
\end{align}
where, $\mathcal{U}_{m}$ is constructed using the states of $m^{\rm th}$ Floquet subspace. Hence, 
$Q_{xy,m}^{(0)}$ takes the form
\begin{align}
Q_{xy,m}^{(0)}=\frac{1}{2\pi} \;\textrm{Im} \;[\textrm{Tr}  \ln \left(\mathcal{U}_m^\dagger \mathcal{W}_m \right)]\ .
\end{align}

Therefore, we calculate the quadrupole moment for the $m^{\rm th}$ Floquet zone as $Q_{xy,m} = Q_{xy,m}^{(0)}-Q_{xy}^{\rm a}$. We find that $Q_{xy,m}$ is always having a quantized value of $0.5$ for all the Floquet zone \ie $m=0,1,2,3$, corroborating the presence of dressed corner modes shifted at quasi-energy $\omega/2$.

%----------------------------------------------------------------------
\section{Summary and Conclusions}{\label{sec:IV}}
%----------------------------------------------------------------------
To summarize, in this article, we have explored the possibility of obtaining both conventional topological insulator (first order) and higher order topological insulator (second order) starting from a 2D trivial semimetal model via Floquet engineering. We have first formulated the Floquet Hamiltonian using Brillouin-Wigner perturbation expansion in the high frequency limit ($\omega \gg t_{1}, t_{2}$) and truncated after $1/\omega$ order. The higher order terms give rise to smaller contributions. As the periodic drive is turned on (circularly polarized light in our case), the time reversal symmetry being broken due to the circular nature of the polarized light and the system becomes an FTI accommodating two counter-propagating edge modes. This topological phase transition is identified by the abrupt change of the Chern number of the quasi-energy bands. With the enhancement of the driving strength, we observe another topological phase transition near $A=3.35$ which occurs with concomitant band gap closing in the bulk. This Floquet topological phase transition from semimetal to FTI is similar to the well known Floquet topological insulator formation by driving pristine Graphene~\cite{usaj2014irradiated,perez2014floquet}.

As the crystalline $\mathcal{C}_4$ symmetry is broken in the square lattice system by adding a Wilson mass term, we realize a static second order topological insulating phase hosting corner modes at the four corners. These in-gap corner modes are demonstrated by both the eigenvalue spectrum as well as LDOS. The static HOTI phase is identified by quantized quadrupolar moment $Q_{xy}^{(0)}$ which always exhibits the value of $0.5$. We numerically compute $Q_{xy}^{(0)}$ which comes out to be $0.5$ within numerical accuracy for our model as soon as the Wilson mass term is added to the Hamiltonian \ie $\mathcal{C}_4$ is broken along with $\mathcal{T}$. When the external drive is turned on, the system becomes Floquet HOTI accommodating Floquet corner modes. For the high frequency drive, we evaluate the Floquet quadrupole moment $Q_{xy}$ which still comes out to be the same quantized value $0.5$ for any strength of the driving amplitude $A$ and consequently Floquet corner modes appear localized at the four corners of the system. We illustrate these corner modes in the LDOS behavior. However, in this frequency regime, it is not apparent to distinguish between the static and dressed corner modes by changing the strength of drive. Therefore, we also explore the HOTI phase in the intermediate frequency regime by truncating the full Floquet Hamiltonian in the subspace of four Floquet sub-bands: $m=0,1,2,3$ and thus taking into consideration real photon transitions within the subbands. We emphasize the appearance of {\it{dressed corner modes}} at quasi-energy 
$\omega/2$ (due to origin shift from zero) from the quasi-energy spectra as well as via calculating LDOS.

In case of high frequency approximation, the Floquet edge and corner modes that appear in respective first order TI and SOTI phase of our model system are still quasi-static as we have formulated our problem based on the effective Hamiltonian picture in that limit. In this limit, only the virtual photon transition has been taken into account. Technically, the full Floquet Hamiltonian in the extended Sambe space is projected back to the zero photon subspace using a high-frequency expansion based on the BW perturbation theory~\cite{mikami2016brillouin}. Thus, although, the obtained zero energy corner modes in presence of the drive might push one to think that the circular polarized light being a kind of perturbation against which the system is robust; we would like to emphasize that the new corner modes are not the same as that of static; rather they are quasi-static Floquet corner mode due to virtual photon transition. 
%in some sense ``dressed corner mode" where ``dressing" emerges due to the coupling of the electrons with the photons. 
In contrary, in case of intermediate frequency regime, real photon transition within the Floquet sub-bands are taken into consideration and as a result, one can realize the dressed corner modes at quasi-energy $\omega/2$ (shifted from zero) of the drive.

Throughout our calculations, we have treated our square lattice model to be disorder free and at zero temperature. Although, this might not be the case in a practical situation. Nevertheless, corner modes should be persistent in the presence of weak disorder and at finite temperature with the disorder and temperature scale being smaller than the bulk bandgap of the system. However, the effect of strong 
disorder with its strength being comparable to the bandwidth and in the presence of external irradiation can be very interesting and is beyond the scope of the present work.

The other models~\cite{rodriguez2019higher, PhysRevB.99.045441} that have been studied so far, need either $\pi$ flux dimerized square lattice with inhomogeneous hopping amplitudes or 1D array of topological insulators (for \eg~Su-Schrieffer-Heeger (SSH) model realized in Polyacetylene) tunnel coupled to each other to realize Floquet higher order topological phases. These models may be more challenging to engineer from experimental point of view due to the requirement of several parameters and coupling of nanowire/Polyacetylene chains. On the other hand, our starting model is a square lattice with homogeneous nearest-neighbour hoppings and such system can possibly be realized in cold atomic systems compared to the others.

As far as practical realization of HOTI is concerned, 2D SOTI has recently been realized in kagome lattice using acoustic measurements~\cite{XueAcousticKagome}, in photonic crystals set up with near field scanning measurement technique~\cite{PhotonicChen,PhotonicXie} and in electrical circuits setup using spectroscopic measurements~\cite{Peterson2018,Imhof2018}. Our 2D model, therefore, may also be possible to engineer either in such systems or in cold atomic systems and can be a platform to understand and discover the Floquet higher order topological phases and Floquet corner modes using local measurements (\eg scanning tunneling microscope (STM)) in presence of periodic drive with circular polarization. 

%---------------------------------------------------------------------------------
\acknowledgments{}
%---------------------------------------------------------------------------------
We acknowledge Adhip Agarwala and Tanay Nag for helpful discussions. We acknowledge SAMKHYA: High Performance Computing Facility provided by Institute of Physics, Bhubaneswar, for our numerical computation.

\bibliography{bibfile}{}

\end{document}